\documentclass[aip,jap,reprint,frontmatterverbose]{revtex4-1}
\usepackage{graphicx}
\usepackage{ulem}
\usepackage{amssymb}
\usepackage{amsmath}
\usepackage{todonotes}
\usepackage{wasysym}
\usepackage{relsize}
\usepackage{bm}
\bibpunct{}{}{,}{s}{}{}
\usepackage{float}
\usepackage{epstopdf}

\begin{document}

%]
\newcommand{\be}{\begin{equation}}
\newcommand{\ee}{\end{equation}}
\newcommand{\bea}{\begin{eqnarray}}
\newcommand{\eea}{\end{eqnarray}}
\newcommand{\bsp}{\begin{split}}
\newcommand{\esp}{\end{split}}
\newcommand{\Tbar}{{\bar{T}}}
\newcommand{\En}{{\cal E}}
\newcommand{\K}{{\cal K}}
\newcommand{\GC}{{\cal \tt G}}
\newcommand{\Lop}{{\cal L}}
\newcommand{\DB}[1]{\marginpar{\footnotesize DB: #1}}
\newcommand{\q}{\vec{q}}
\newcommand{\kt}{\tilde{k}}
\newcommand{\Lopn}{\tilde{\Lop}}
\newcommand{\noi}{\noindent}
\newcommand{\ovn}{\bar{n}}
\newcommand{\ovx}{\bar{x}}
\newcommand{\ovE}{\bar{E}}
\newcommand{\ovV}{\bar{V}}
\newcommand{\ovU}{\bar{U}}
\newcommand{\ovJ}{\bar{J}}
\newcommand{\calE}{{\cal E}}
\newcommand{\ovphi}{\bar{\phi}}
\newcommand{\zt}{\tilde{z}}
\newcommand{\rt}{\tilde{\rho}}
\newcommand{\tth}{\tilde{\theta}}
\newcommand{\nuv}{{\rm v}}
\newcommand{\ck}{{\cal K}}
\newcommand{\cc}{{\cal C}}
\newcommand{\ca}{{\cal A}}
\newcommand{\cb}{{\cal B}}
\newcommand{\cg}{{\cal G}}
\newcommand{\ce}{{\cal E}}
\newcommand{\cn}{{\cal N}}
\newcommand{\fn}{{\small {\rm  FN}}}
\newcommand\norm[1]{\left\lVert#1\right\rVert}
\newcommand{\Afn}{A_{\text{FN}}}
\newcommand{\Bfn}{B_{\text{FN}}}
\newcommand{\mg}{\text{MG}}
\newcommand{\sh}{\text{SF}}
\newcommand{\tcc}{\text{cc}}
\newcommand{\tcm}{\text{cm}}
\newcommand{\tma}{\text{ma}}
\newcommand{\tmax}{\text{0}}
%\newcommand{\tth}{\tilde{\theta}}
%\newpage
%\noindent

\title{Fast and accurate determination of the curvature-corrected field emission current}

\author{Debabrata Biswas}\email{dbiswas.hbni@gmail.com}
\author{Rajasree Ramachandran}
\affiliation{
Bhabha Atomic Research Centre,
Mumbai 400 085, INDIA}
\affiliation{
  Homi Bhabha National Institute, Mumbai 400 094, INDIA}

\begin{abstract}
  The curvature-corrected field emission current density, obtained by linearizing at or below the
  Fermi energy, is investigated. Two special cases, corresponding to the peak of the normal energy
  distribution and the mean normal energy, are considered. It is found that the current density
  evaluated using the mean normal energy results in errors in the net emission current
  below 3\% for apex radius of curvature,
  $R_a \geq 5$nm and for apex fields $E_a$ in the range $3-10$~V/nm for an emitter
  having work-function $\phi = 4.5$eV.
  An analytical expression for the net field emission current is also obtained
  for locally parabolic tips using the generalized
  cosine law. The errors are found to be below 6\% for $R_a \geq 5$nm over an identical range
  of apex field strengths. The benchmark current is obtained by numerically integrating
  the current density over the emitter surface and the current density itself computed by
  integrating over the energy states using the exact Gamow factor and the Kemble form
  for the WKB transmission coefficient. The analytical expression results
  in a remarkable speed-up in the computation of the net emission current and is
  especially useful for large area field emitters having tens of thousands of emission sites.
  \end{abstract}

\maketitle

\section{Introduction}
\label{sec:intro}

Recent studies have shown that field emitters with tip radius in the nanometer range
can be best modelled accurately by taking into account the variation in local field in the
tunneling region, which is roughly 1-2nm from the emitter surface depending on the
field strength\cite{db_rk_2019,lee2018,db_rr_2019,rr_db_2021,db_rr_2021}.
When the apex radius of curvature ($R_a$) of the emitter is large ($R_a > 100$nm),
the local field is roughly constant in this region even though the field enhancement
factor itself may be large\cite{db_rr_2019}. Thus, the Murphy-Good current
density\cite{murphy,forbes2006,FD2007,DF2008,jensen_book,jensen_rTF,forbes2019,db_dist,db_gamow}
is quite likely adequate\cite{db_rr_2019} for $R_a > 100$nm while for emitters with $R_a < 100$nm,
errors first start building up at smaller field strengths and for $R_a \leq 10$nm,
the errors become large over a wide range of fields\cite{db_rr_2019,db_rr_2021}.

The necessity for curvature-corrections was illustrated recently\cite{db_rk_2019}
using the experimental results for a single Molybdenum emitter tip\cite{lee2018}
with a FESEM-estimated endcap apex radius of curvature
in the 5-10nm range with the square-shaped pyramidal base having a side-length $L_b \in [1.25,1.35]\mu$m.
Interestingly, even on using the Fowler-Nordheim\cite{FN} current density that ignores image-charge
contribution and seriously under-predicts the current density, the fit
was good\cite{db_rk_2019,assumed_constant}
but required an emission area  of $130 000 \text{nm}^2$.
In contrast, the area of a hemisphere of radius 10nm is only
about $628 \text{nm}^2$. On the other hand\cite{db_rk_2019}, the Murphy-Good current density (that
takes into account image-charge contribution to the tunneling potential \cite{Nordheim}), used
with the generalized cosine law\cite{db_ultram,physE} of local field variation around the emitter tip, had
a best fit to the experimental data with $R_a = 9.79$nm which is within the estimated range of $R_a$.
However, the value of field enhancement required the base-length $L_b$ to be $0.65\mu$m which
is clearly outside
the $1.25-1.35\mu$m range. Thus, while the Fowler-Nordheim current density has gross non-conformity
with the physical dimensions, the Murphy-Good current density seems to be in need of a correction.
Indeed, on using a curvature-corrected (CC) expression for emission current\cite{db_rr_2019}, the best fit to
experimental data required\cite{db_rk_2019} $R_a \approx 5.41$nm and $L_b \approx 1.275\mu$m, both of
which are within the range of their respective estimated values. This one-off validation could be a
coincidence and more such experiments, observation and data analysis are required to explore and
put on a firm footing, the limits of validity of each model \cite{see1,popov23}.

The evidence so far seems to suggest that a curvature-corrected field emission theory is
necessary for nano-tipped emitters.
An elementary form of this\cite{db_rr_2019} was used in Ref.~[\onlinecite{db_rk_2019}],
based on a tunneling potential having a single correction term. Since then, an approximately
universal tunneling potential having an additional curvature correction term has been
established\cite{db_gs_rk_2016}
using the nonlinear line charge model \cite{mesa,pogorelov,harris2015,harris2016,db_gs_rk_2016}
and tested against the finite-element
software COMSOL\cite{rr_db_2021}. A curvature-corrected analytical current density
has also been determined\cite{db_rr_2021} by suitably algebraic approximation of
the exact Gamow factor and its linearization at the Fermi energy.
While the results are promising, there is a scope for improving its
accuracy by choosing a different linearization-energy. It is also desirable to have
an  analytical expression for the net field emission current applicable for
$R_a \geq 5$nm over a wide range of fields. The present communication seeks to establish 
accurate analytical expressions for both, the curvature-corrected local current density, as well as the
net emission current for smooth locally parabolic emitters.

The issue of accuracy in analytical expressions for current density
has recently been investigated in Ref. [\onlinecite{db_gamow}] for emitters where curvature
corrections are unimportant ($R_a > 100$nm). The three major factors investigated were:
(a) the form in which the Gamow factor, $G$, is cast (b) the use of $e^{-G}$ to determine the
transmission coefficient and (c) the energy at which the Gamow factor should be linearized
in order to obtain an approximate analytical form for the current density. It was
found\cite{db_gamow} that if an analytical form of the current
density is used to determine the net emission current, only the second and third factors
are important.
For instance, the use of $e^{-G}$ to determine the
transmission coefficient leads to errors at larger local fields where the tunneling barrier
transitions from `strong' to `weak'. A better way of determining the transmission coefficient
within the WKB approximation is the Kemble\cite{kemble,forbes2008pre} formula $(1 + e^G)^{-1}$.
Another significant cause of error can be ascribed to the energy at which the Gamow
factor is linearized
in order to obtain an approximate analytical form for the current density.
In the traditional approach to cold field emission, the Gamow factor is linearized
at the Fermi energy. While this holds at smaller values of the local field, it leads to
large errors at higher fields due to the shift in the normal energy distribution away from
the Fermi energy.  In the following, we shall continue to use the traditional representation of the
Gamow factor in term of the Forbes approximation\cite{forbes2006} for the WKB integral, 
and add curvature corrections to it.

In section \ref{sec:newexpansion}, we shall make use of a curvature-corrected
current density that makes use of a Kemble correction and a shifted point
of linearization. We shall compare the results by choosing the energy corresponding to
the peak of the normal energy distribution as well as the mean normal energy.
While both results are encouraging, the mean normal energy is more
accurate especially at lower field strengths. Finally, an
evaluation of the net field-emission current is carried out
using the generalized cosine law in section \ref{sec:newcurrent} and compared
with the exact WKB result. Summary and discussions form the concluding section.

\section{An accurate curvature-corrected current density}
\label{sec:newexpansion}
A widely adopted method to obtain an
analytical expression for the current density is to Taylor expand the 
Gamow factor about the Fermi energy $\ce_F$ in order to carry out the energy integration.
Recent studies\cite{db_gamow} show that this is
adequate at smaller local field strengths for which the electrons closer to the Fermi 
energy predominantly tunnel through. As the field strength increases, the height and width of 
the tunneling barrier decreases and the electrons well below the Fermi energy 
start contributing to the net emitted current. This is evident from the shift in the 
 peak of the normal energy distribution\cite{db_dist} of the emitted electrons as the local field 
 increases. Hence, for cold field emission, an expansion of the Gamow factor around
 the peak of the normal energy distribution or the mean normal energy
 seems preferable. This is likely to yield a better approximation for
 field emission current density applicable over a wide range of fields.

 The use of $e^{-G}$ is also a factor that contributes to the errors at higher
 fields where the barrier becomes weak. The transmission coefficient in the
 Kemble form\cite{kemble} can be approximated as\cite{db_gamow}

 \be
 T(\ce) = \frac{1}{1 + e^G} \approx e^{-G} \left[ 1 - e^{-G} \right]. \label{eq:TC}
 \ee

 \noi
 Used alongside the linearization of the Gamow factor, this is likely to provide
 a simple yet reasonably accurate expression for the field emission current density.

 \subsection{Expansion of the Gamow factor and the curvature corrected current density}

 The Gamow factor is expressed as 
 
\be
G = g \int_{s_1}^{s_2} \sqrt{V_T(s)-\ce}  ds .\label{eq:Gamow}
\ee

\noi
Here, $g = 2 \sqrt{2 m}/\hbar$, $m$ the mass of the electron and $\hbar$ the 
reduced Planck's constant $h/(2\pi)$. In Eq.~(\ref{eq:Gamow}), $V_T$ is the tunneling potential energy,
$\ce$ is the normal component of electron energy at the emission surface and $s_1, s_2$ are the 
zeroes of the integrand. The curvature-corrected form of
the tunneling potential energy is \cite{db_rr_gs_2018,rr_db_2021} 

\be
V_T(s) \approx \ce_F + \phi+V_{ext}(s)-\frac{B}{s(1+s/2R)} \label{eq:tunnelpot}
\ee
where, $\phi$ is the work function, $\ce_F$ the Fermi energy while
the external potential energy $V_{ext}$ takes the form,
\be   
V_{ext}(s) \approx -q E_l s \Big[1-\frac{s}{R}+\frac{4}{3}\big(\frac{s}{R}\big)^2\Big] \label{eq:ext_pot}
\ee
with $q$ the magnitude of electronic charge, $B = q^2/(16 \pi \epsilon_0)$, $E_l$ the local electric field,  and $s$ denoting the normal distance from the surface of the emitter. The quantity $R^{-1}$ is the mean curvature\cite{spherical,ludwick2021} so that $R$ is the harmonic mean of the principle radii of curvature $R_1$ and $R_2$ at the emission site i.e. $R = 2/(R_1^{-1} + R_2^{-1})$. The curvature-corrected external potential of Eq.~\ref{eq:ext_pot} follows directly from Eq.~(35) of Ref. [\onlinecite{rr_db_2021}] which holds in the region close to the apex for all axially symmetric emitters in a parallel plate diode configuration.
For a more detailed exposition, the reader may refer to appendix \ref{appendixA} on the tunneling potential\cite{R2}.

Using the curvature-corrected tunneling potential energy of Eq.~(\ref{eq:tunnelpot}), an approximate 
form for the Gamow factor can be found numerically to be\cite{db_rr_2021}
\bea
G &= &\frac{2}{3} g \frac{\varphi^{3/2}}{q E_l}[\nu(y)+x w_1(y)+x^2 w_2(y)+x^3 w_3(y)] \label{eq:G3} \\
 & =& \frac{2}{3} g \frac{\varphi^{3/2}}{q E_l} \nu_c(y) .
\eea 

\noi
Here, $\varphi = \ce_F + \phi-\ce$, $x = \varphi/(qE_lR)$, $y = 2\sqrt{qBE_l}/\varphi$ and the 
curvature-corrected barrier function $\nu_c(y)=\nu(y)+x w_1(y)+x^2 w_2(y)+x^3 w_3(y)$,
where\cite{rr_db_2021}

\bea
\nu(y) & = &   1 - y^2 + \frac{1}{3} y^2 \ln y \\ 
w_1(y) & = &\frac{10}{13} - \frac{2}{11}y^2 + \frac{1}{80}y^4 + \frac{1}{200}y^2 \ln y    \\
w_2(y) & =  & \frac{10}{11} + \frac{2}{11}y^2 - \frac{1}{6}y^4 + \frac{1}{200}y^2 \ln y  \\ 
w_3(y) & =  & -\frac{41}{10} + \frac{39}{20}y^2 + \frac{1}{3}y^4 - \frac{1}{150}y^2 \ln y . \label{eq:ws}
\eea

\noi
Note that $x w_1(y), x^2 w_2(y), x^3 w_3(y)$ are the curvature corrections that arise
due to $R$ dependent terms in the external as well as image charge potential.
As $R \to \infty$ in the planar limit, $x \to 0$,
so that $\nu_c(y)$ reduces to $\nu(y)$ which corresponds to the use of the Schottky-Nordheim 
barrier.

We shall hereafter denote the linearization energy by $\ce_m$. On expansion of the curvature-corrected Gamow
factor and retaining the linear term, we obtain

\be
G(\ce) \approx G(\ce_m) - (\ce - \ce_m) \frac{t_\tcm}{d_m}
\ee

\noi
where $t_\tcm = t_c(\ce_m)$ with

\be
\begin{split}
  t_\tcm & = t(y_m) + x_m t_1(y_m) + x_m^2 t_2(y_m) + x_m^3 t_3(y_m) \\
  & = \left( 1 + \frac{y_m^2}{9} - \frac{1}{9}y_m^2 \ln y_m \right) +  \\
& x_m \left(\frac{25}{13} - \frac{237}{1100} y_m^2 - \frac{1}{480}y_m^4 - \frac{7}{1200} y_m^2 \ln y_m \right) +  \\
& x_m^2 \left( \frac{70}{33} + \frac{589}{3300} y_m^2 + \frac{1}{18}y_m^4 + \frac{1}{200} y_m^2 \ln y_m \right) + \\
& x_m^3 \left( -\frac{123}{10} + \frac{2929}{900} y_m^2 + \frac{1}{9}y_m^4 - \frac{1}{90} y_m^2 \ln y_m \right)
\end{split}  \label{eq:tc}
\ee

\noi
where $y_m  =  c_s \sqrt{E_l}/\varphi_m$, $d_m^{-1}  = g \frac{\varphi_m^{1/2}}{E_l}$, $c_s = 1.199985~ {\rm eV}~({\rm V/nm})^{-1/2} $,
$\varphi_m = \ce_F + \phi - \ce_m$ and $x_m = \varphi_m/(qE_lR)$. The Gamow factor
at $\ce_m$ can be expressed as

\be
G(\ce_m) = \Bfn \varphi_m^{3/2} \frac{\nu_\tcm}{E_l} \label{eq:Gm}
\ee

\noi
where $\nu_\tcm = \nu(y_m)+x_m w_1(y_m)+x_m^2 w_2(y_m)+x_m^3 w_3(y_m)$.
The field emission current density

\bea
J & = & \frac{2mq}{(2\pi)^2 \hbar^3} \int_{0}^{\calE_F} (\ce_F - \ce)\frac{1}{1 + e^{G(\ce)}}~d{\ce}  \\
& \approx & \frac{2mq}{(2\pi)^2 \hbar^3} \int_{0}^{\calE_F} (\ce_F - \ce) e^{-G(\ce)} \left[ 1 - e^{-G(\ce)} + \ldots \right]~d{\ce}   \nonumber
\eea

\noi
can be expressed on completing the integration over energy states as

\bea
J_{\tcc}^m & \approx & \Afn  \frac{1}{\varphi_m}\frac{E_l^2}{t_\tcm^2} e^{-\cb_\tcc} \left( 1 - \frac{e^{-\cb_\tcc}}{4} \right) \label{eq:MGshift} \\
\cb_\tcc & = & \Bfn \varphi_m^{3/2} \frac{\nu_\tcm}{E_l} - \frac{t_\tcm}{d_m}(\ce_F - \ce_m)  \label{eq:newB}  
\eea

\noi
where $A_\fn~\simeq~1.541434~{\rm \mu A~eV~V}^{-2}$,
$B_\fn~\simeq 6.830890~{\rm eV}^{-3/2}~{\rm V~nm}^{-1}$ are the usual Fowler-Nordheim
constants. The curvature-corrected current density $J_\tcc^{m}$ (Eq.~(\ref{eq:MGshift})), with the
incorporation of the first Kemble correction and linearization of the Gamow factor at $\ce_m$
provides an analytical expression that can be used to evaluate the net field
emission current from a curved emitter, either by numerically integrating over the surface
or by using the local field variation over the emitter surface to obtain an approximate
analytical expression for the net field emission current.

\subsection{Numerical verification}
\label{sec:num}

The exact WKB result (referred to hereafter as the benchmark) obtained by (i) finding the Gamow
factor exactly by numerical integration (ii) use of the Kemble form of transmission coefficient
and (iii) numerical integration over energy to obtain the current
density, can be used to validate Eq.~(\ref{eq:MGshift}).  Since we shall be comparing net emission
currents rather than current-densities,
the local current density is integrated over the surface near the apex to obtain the
net current numerically.

The geometrical entity we are focusing on is an axially-symmetric emitter having
an apex radius of curvature $R_a$ and height $h = 300 R_a$. It is mounted on a parallel
plate diode where the generalized cosine law\cite{db_ultram,physE} of local field variation holds:

\be
E_l = E_a \frac{z/h}{\sqrt{(z/h)^2 + (\rho/R_a)^2}} = E_a \cos\tth \label{eq:cos}.
\ee

\noi
In the above, $h$ is the height of the emitter, $R_a$ is the apex radius of curvature
and $E_a$ the apex field. Eq.~(\ref{eq:cos}) holds for all axially symmetric emitters
where the tips are locally approximated well by a parabola $z \approx h - \rho^2/(2R_a)$
upto $\rho \approx R_a$. Thus the only parameters required are $h, R_a$ and
the apex field\cite{enhancement,db_2018a,db_anode,review2022} $E_a$, since the
generalized cosine law\cite{db_ultram,physE} for local fields holds for such emitter-tips.
Note that the benchmark also uses the parabolic approximation and the generalized cosine law
for determining the net emission current\cite{high_field}. In the following, we shall
consider $\ce_F = 8.5$eV and $\phi = 4.5$eV. The apex fields considered are in the
range [3,10]~V/nm which correspond to scaled barrier fields\cite{forbes2008} $E_a/E_\phi$ in the range
0.21333 - 0.71109 where $E_\phi = (0.6944617~ \text{eV}^{-2} \text{V} \text{nm}^{-1}) \phi^2$.

\begin{figure}[thb]
\vskip -0.5 cm
%\centering
\hspace*{-01.05cm}\includegraphics[width=0.62\textwidth]{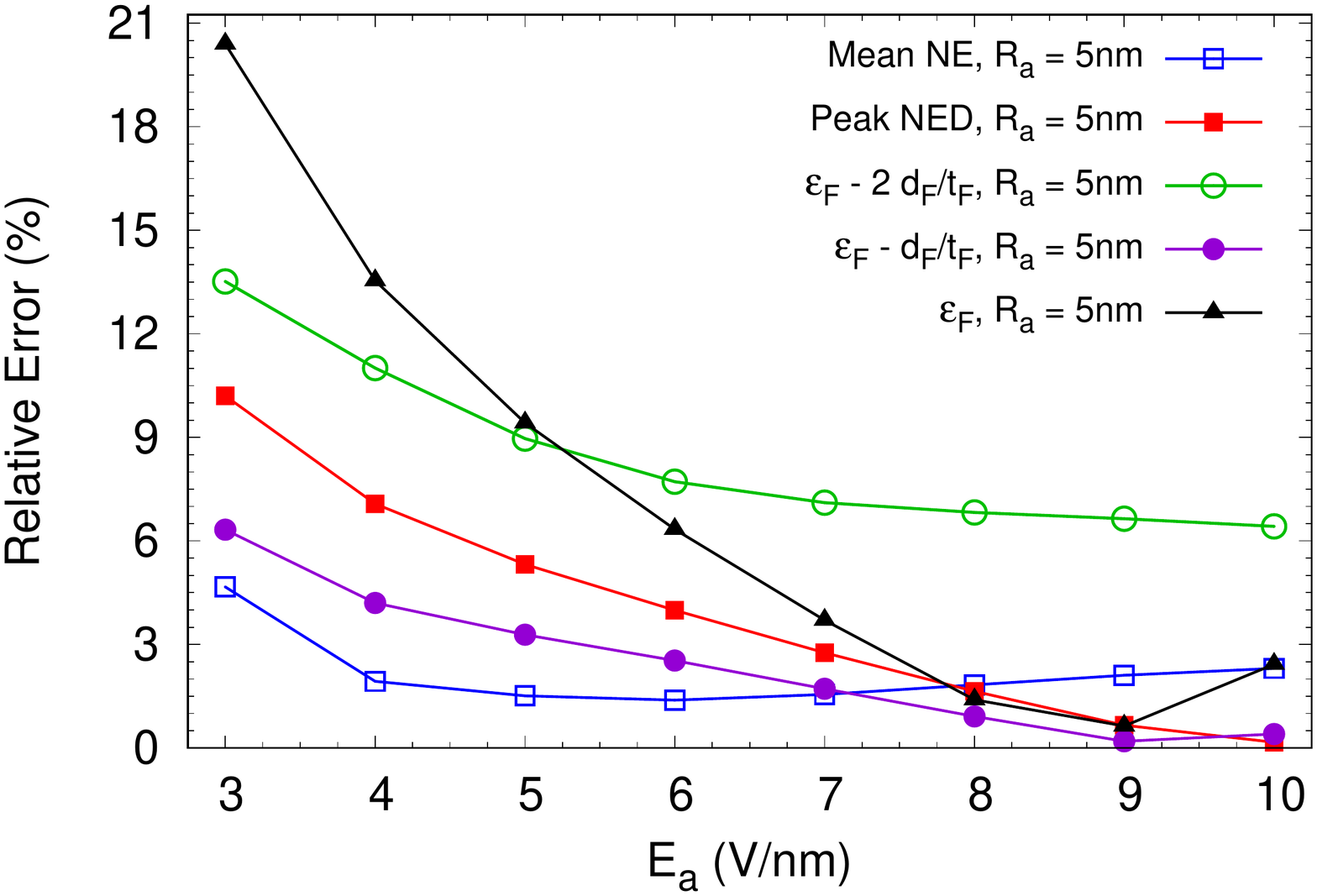}
\vskip -0.75 cm
\caption{The absolute relative error in the net emission current with respect to the exact
  WKB result. Five cases are shown with various linearization energy $\ce_m$.
  `Mean NE' refers to the exact mean normal energy, `Peak NED' refers to the exact energy
  at which the normal energy distribution peaks, `$\ce_F$' refers to $\ce_m = \ce_F$,
  `$\ce_F - 2d_F/t_F$' is the approximate mean normal energy, while `$\ce_F - d_F/t_F$' is the
  approximate peak of the normal energy distribution. 
  }
\label{fig:5nm}
\end{figure}

\begin{figure}[thb]
\vskip -0.5 cm
%\centering
\hspace*{-1.05cm}\includegraphics[width=0.62\textwidth]{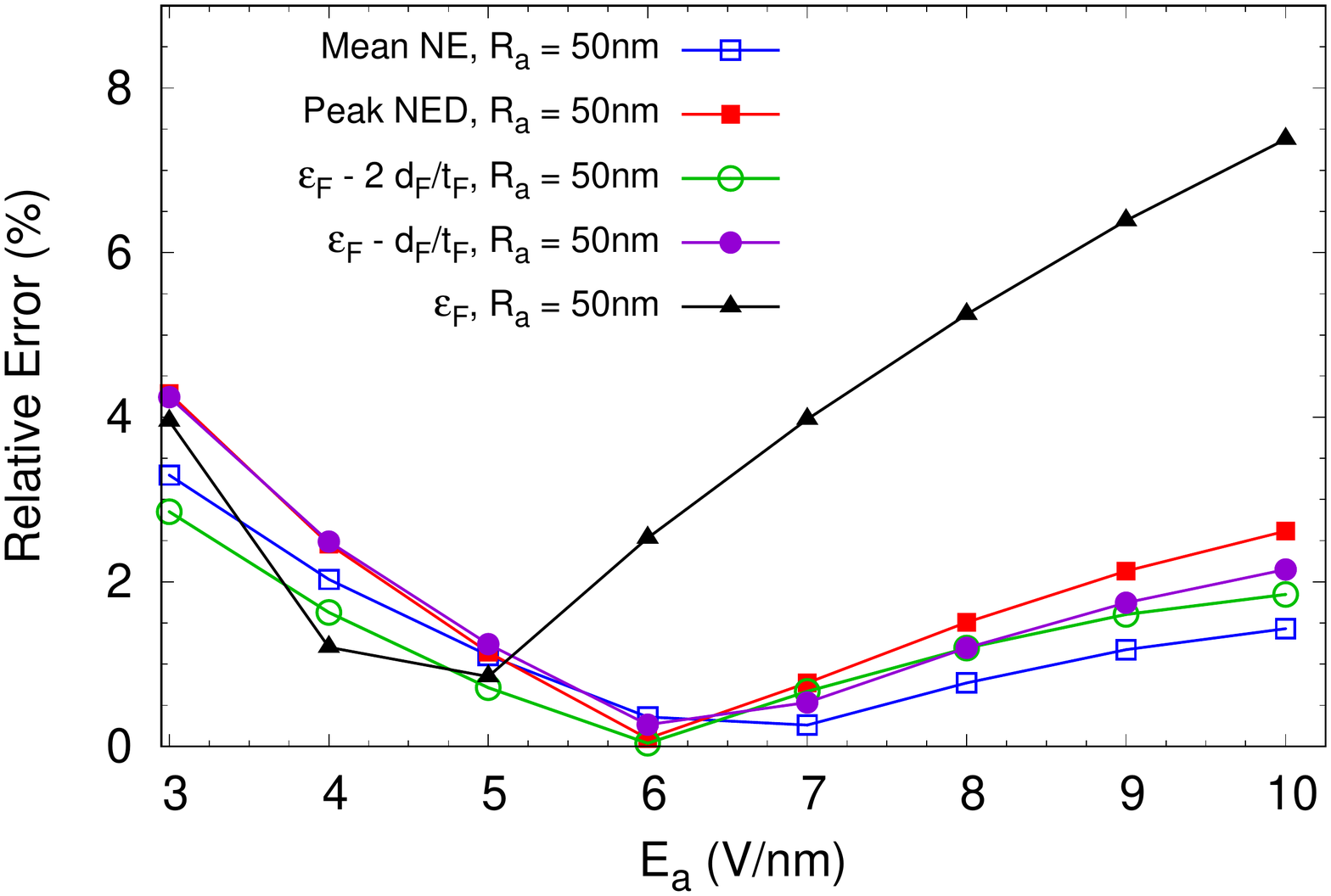}
\vskip -0.75 cm
\caption{As in case of Fig.~\ref{fig:5nm} with $R_a = 50$nm. Note that the error
for $\ce_m = \ce_F$ increases at higher values of $E_a$.}
\label{fig:50nm}
\end{figure}

It is clear that there are severals factors at play when comparing the error with respect to the
exact WKB result. We shall discuss two of these from the broad picture available to us.
The first is the effect of curvature correction which reflects in the
approximate Gamow factor in Eq.~(\ref{eq:G3}). Since the expansion is in powers of
$x = \varphi/(qE_lR)$, the approximate Gamow factor is prone to errors at smaller values of
$E_l$ and $R$. Thus, irrespective of the energy at which the linearization is carried
out, lower fields and radius of curvature are prone to errors. 
In general, at higher $R$ and $E_l$, the curvature errors are expected to reduce.
The second important consideration is the energy at which the linearization is carried out.
Since the peak of the normal energy distribution moves away from $\ce_F$ at higher fields for a given
$R_a$, linearization at $\ce_m = \ce_F$ should in general lead to larger errors
at higher fields strengths. Apart from these two, there are other subtle effects that decide
the magnitude of relative error at a given field strength as we shall see. Note that on the
surface of an emitter, $E_l$ reduces away from the apex while $R$ increases and this leads
to a mild decrease in the expansion parameter $x$.

\begin{figure}[thb]
\vskip -0.5 cm
%\centering
\hspace*{-1.05cm}\includegraphics[width=0.62\textwidth]{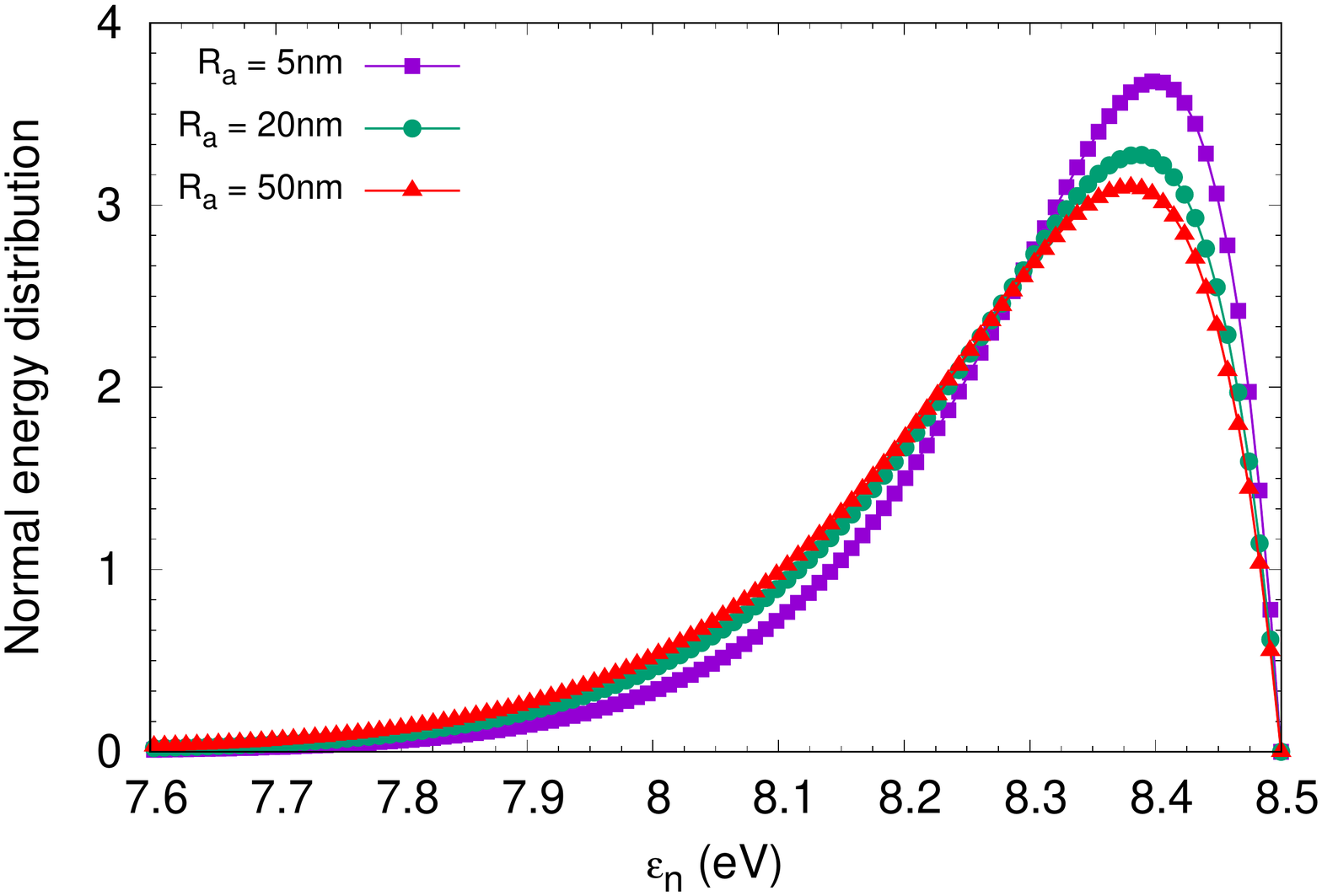}
\vskip -0.75 cm
\caption{The normal energy distribution at $E_a = 3$V/nm for $R_a = 5,~20~ \text{and}~ 50$nm. Note the shift in the
  distribution away from the $\ce_F$ (= 8.5eV here) for larger values of $R_a$.}
\label{fig:ned}
\end{figure}

\begin{figure}[thb]
\vskip -0.5 cm
%\centering
\hspace*{-1.05cm}\includegraphics[width=0.6\textwidth]{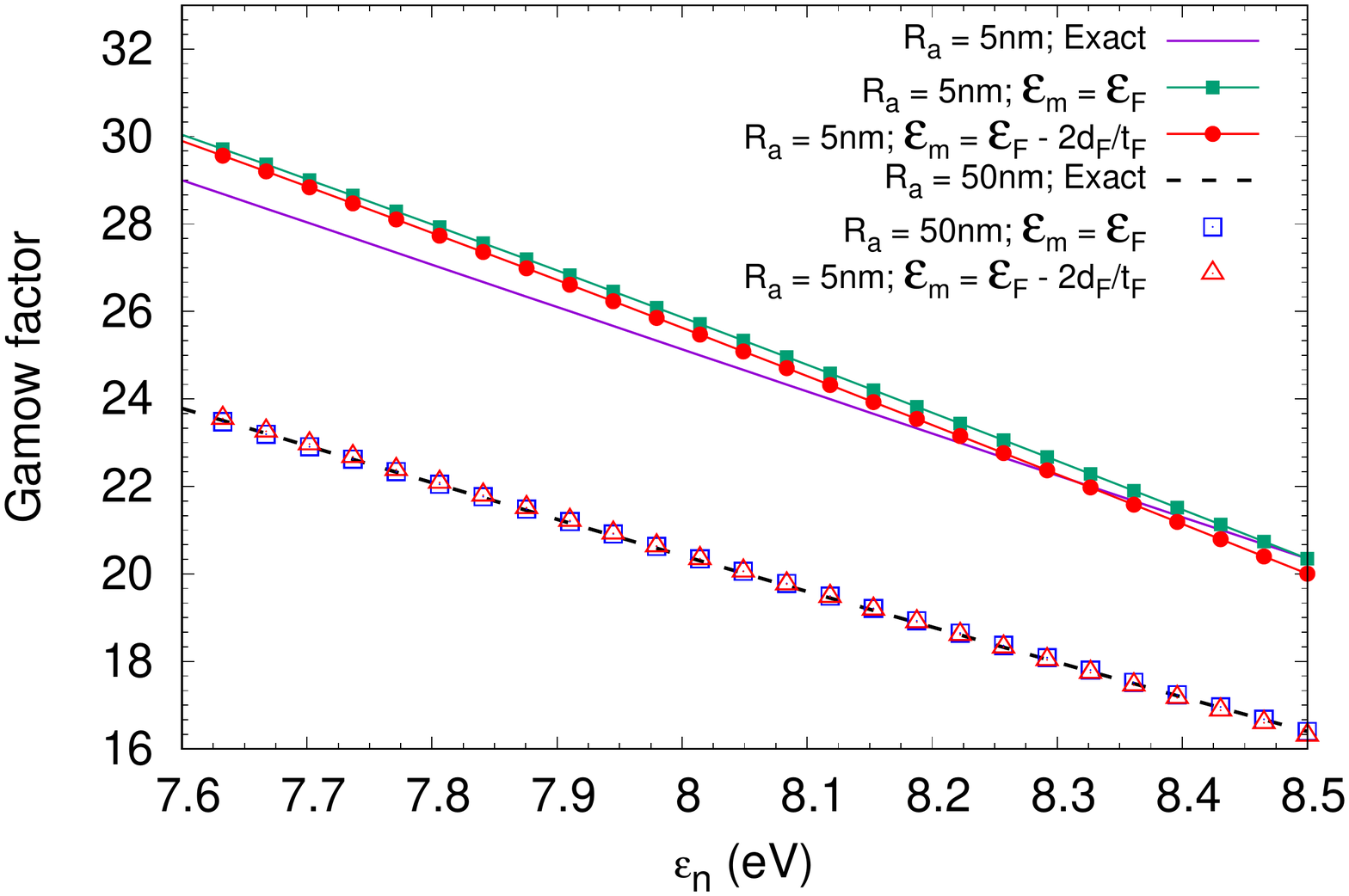}
\vskip -0.75 cm
\caption{The exact Gamow factor at $E_a = 3$V/nm is compared with the linearized
  Gamow factors with the point of linearization at
  $\ce_m = \ce_F$ and $\ce_m = \ce_F - 2d_F/t_F$. The upper set of curves correspond to $R_a = 5$nm while
  the lower set is for $R_a = 50$nm. The two linearized versions are nearly equivalent at $R_a = 50$nm
  while, for $R_a = 5$nm linearizing at the approximate mean energy yields results closer to the exact Gamow
  factor over the energy range of interest.}
\label{fig:gamow}
\end{figure}

With this perspective, we shall compare the absolute relative
errors at $R_a = 5$nm and $R_a = 50$nm shown in Figs.~\ref{fig:5nm}
and \ref{fig:50nm} respectively, for various values of $\ce_m$ displayed in the
legends. Clearly `Mean NE', which refers to the exact mean normal energy  determined
numerically (see appendix \ref{sec:appenMEAN}), performs well at $R_a = 5$nm at all field strengths while $\ce_m = \ce_F$
shows large errors especially at lower fields. Even at $R_a = 50$nm where curvature errors
are expected to be smaller, `Mean NE' 
as well as the approximate mean normal energy ($\ce_m \approx \ce_F - 2d_F/t_F$)
perform well while in case of $\ce = \ce_F$, the linearization error dominates
leading to larger errors at higher field strengths.
The energy value corresponding to the peak of the normal energy distribution (`Peak NED') also gives good results
though the errors are somewhat high for smaller apex fields at $R_a = 5$nm.

Some of the trends in Figs.~\ref{fig:5nm} and \ref{fig:50nm} are easy to understand. For instance, at
$R_a = 5$nm, the errors fall as expected with an increase in $E_a$ in all cases (except for a mild
increase at $\ce_m = \ce_F$ for $E_a > 9$V/nm). The larger than expected error (approximately $21\%$) for
$\ce_m = \ce_F$ at $E_a = 3$V/nm
however seems intriguing. To understand this better, the normal energy distribution (see Fig.~\ref{fig:ned})
for different values of $R_a$ at $E_a = 3$V/nm is quiet instructive. The peak of the normal
energy distribution shifts slightly away from $\ce_F$ as $R_a$ increases. Note also that the distributions have a
long tail. The linearized Gamow factor in the corresponding normal energy range is shown in Fig.~\ref{fig:gamow}.
For $R_a = 5$nm, linearization at
$\ce_F$ results in larger deviations from the exact Gamow factor compared to linearization at $\ce_F - 2d_F/t_F$.
Not surprisingly, the relative error in net emission current drops from about $21\%$ to about $13\%$
in moving from $\ce_m = \ce_F$ to $\ce_m = \ce_F - 2d_F/t_F$.

At $R_a = 50$nm, curvature effects are smaller and the linearized Gamow factor does not noticeably deviate from
the exact Gamow factor (see Fig.~\ref{fig:gamow} for $E_a = 3$V/nm). Thus, the errors remain more or less similar
at all linearization energies. The magnitude of the error at a particular $E_a$ depends on how closely the
linearized Gamow factor approximates the exact Gamow factor over the relevant range of normal energies.
For $E_a > 5$V/nm at $\ce_m = \ce_F$, the increase in error is expected due to the shift in normal energy
distribution away from $\ce_F$ and the corresponding deviation of the linearized Gamow factor from the
exact Gamow factor.

\begin{figure}[thb]
\vskip -0.5 cm
%\centering
\hspace*{-0.950cm}\includegraphics[width=0.6\textwidth]{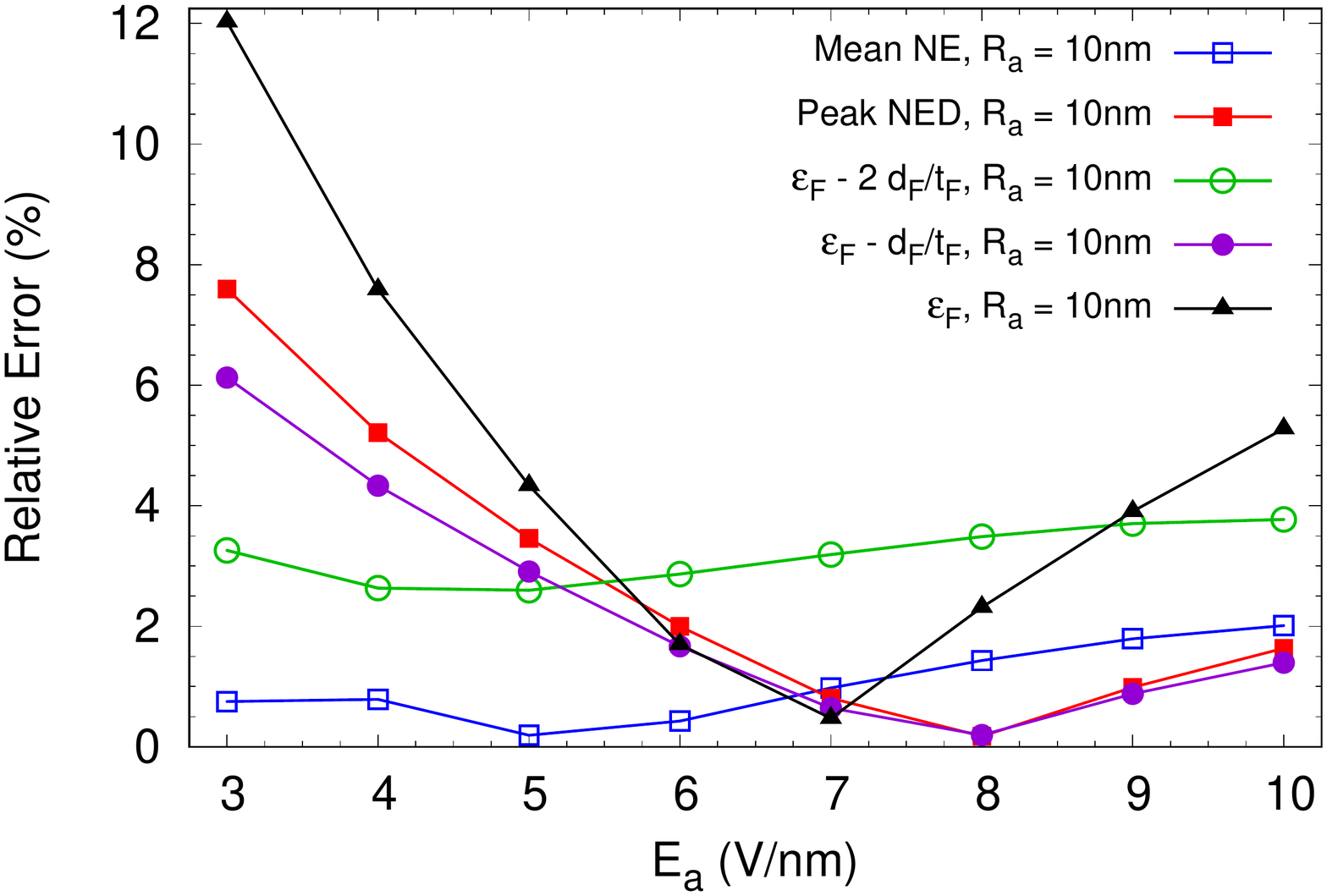}
\vskip -0.75 cm
\caption{As in case of Fig.~\ref{fig:5nm} with $R_a = 10$nm.}
\label{fig:10nm}
\end{figure}

\begin{figure}[thb]
\vskip -0.5 cm
%\centering
\hspace*{-0.95cm}\includegraphics[width=0.6\textwidth]{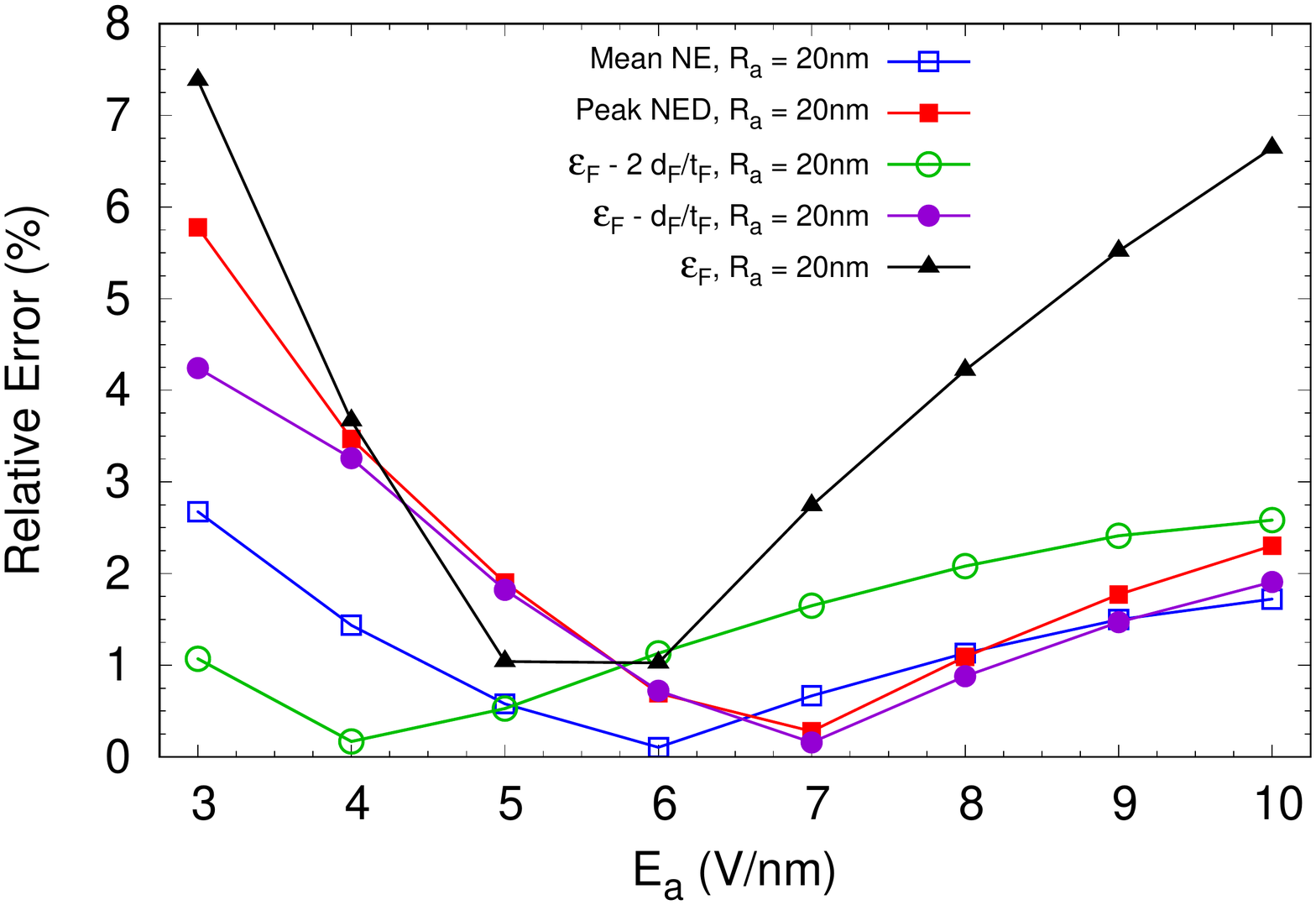}
\vskip -0.75 cm
\caption{As in case of Fig.~\ref{fig:5nm} with $R_a = 20$nm.}
\label{fig:20nm}
\end{figure}

In order to verify that the trend observed in moving from $R_a = 5$nm to $R_a = 50$nm is gradual,
we show the results for $R_a = 10$nm and
$R_a = 20$nm in Figs. \ref{fig:10nm} and \ref{fig:20nm}. It is apparent from
these results that linearization at the exact mean normal energy (`Mean NE') is
optimum for all values of $R_a$ and $E_a$ with errors generally below $3\%$.
The approximate mean normal energy $\ce_m \approx \ce_F - 2d_F/t_F$ is only marginally worse
with errors exceeding 6\% only at $R_a = 5$nm.

\section{The net curvature-corrected emission current}
\label{sec:newcurrent}

The curvature-corrected expression for the current density, with linearization at
the mean normal energy, can be used to arrive at an analytical expression for the
net emission current  on using the generalized cosine law of local field
variation $E_l = E_a \cos\tth$ (Eq.~(\ref{eq:cos})).
Assuming a sharp locally parabolic emitter tip, the total emitted current can be 
evaluated using the expression\cite{db_dist}

\be
I \approx 2\pi R_a^2\int J_\tcc(\tth) \frac{\sin\tth}{\cos^4\tth}\times \cc(\tth)~ d\tth
\ee

\noi
where $\cc(\tth)$ is a correction factor which, for a sharp emitter  ($h/R_a >> 1$),
is approximately unity. In the following we shall assume the emitter to be reasonably
sharp so that $\cc \approx 1$.

The basic idea is to express $J_\tcc$ in terms of $\tth$ by replacing all the local
fields using $E_l = E_a \cos\tth$. A further simplification can be made by the substitution
$1/\cos\tth = 1 + u$ and retaining only terms upto ${\mathcal{O}(u^2)}$ in $\cb_\tcc$
and $t_\tcm^{-2}$. The approximation is expected to be good at lower apex fields
since the emission is limited to an area closer to the apex, while at higher
fields, where the emission area is larger, this might lead to larger errors.

Writing $\cb_\tcc \approx D_0 + D_1 u$ and  $t_\tcm^{-2} \approx F_0 + F_1 u $,
the integration can be carried out easily. Note that, it generally suffices to
integrate upto $\rho = R_a$ which, for a sharp emitter, corresponds to $\tth = \pi/4$
or $ u = \sqrt{2} - 1 = u_\tmax$. Thus,

\be
I  \approx  2\pi R_a^2 \cg \Afn  \frac{1}{\varphi_{\tma}} E_a^2 F_0  e^{-D_0} \label{eq:anI}
\ee

\noi
where $\ce_m$ is the mean normal energy, while

\be
\begin{split}
  \cg  \approx  & \frac{1}{D_1}  + \frac{F_1}{F_0}\frac{1}{D_1^2} - \frac{e^{-D_0}}{4}\left(\frac{1}{2D_1} + \frac{F_1}{F_0}\frac{1}{4D_1^2} \right) - \\
  & e^{-D_1 u_\tmax} \left[  \frac{1}{D_1} + \frac{F_1}{F_0}\frac{1 + D_1 u_\tmax}{D_1^2} -
    \frac{e^{-D_0 - D_1 u_\tmax}}{4} \times \right. \\
  & \left.~~~~~~~~~~ \left(\frac{1}{2D_1} + \frac{F_1}{F_0}\frac{1 + 2D_1 u_\tmax}{4D_1^2} \right)\right].
  \end{split} \label{eq:gfact}
\ee

\noi
Expressions for $D_0, D_1, F_0$ and $F_1$ can be found in appendix \ref{sec:appen}.

\begin{figure}[thb]
\vskip -0.5 cm
%\centering
\hspace*{-1.05cm}\includegraphics[width=0.62\textwidth]{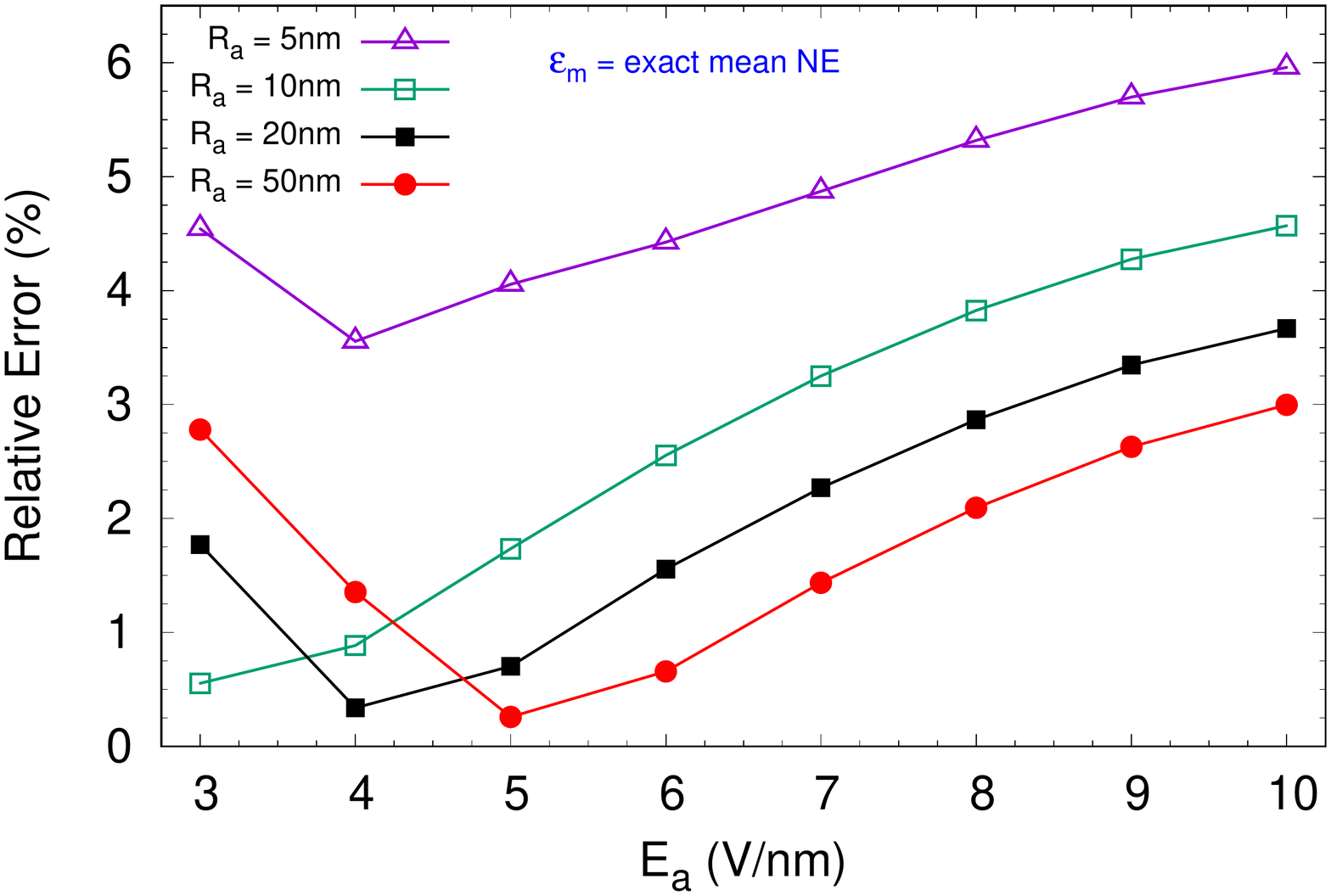}
\vskip -0.75 cm
\caption{The magnitude of the relative error in the analytical expression for
  the curvature-corrected current (Eq.~\ref{eq:anI}) compared to the exact WKB result.
Here $\ce_m$ is the exact mean normal energy.}
\label{fig:D0D1WKB}
\end{figure}

\begin{figure}[thb]
\vskip -0.5 cm
%\centering
\hspace*{-1.05cm}\includegraphics[width=0.62\textwidth]{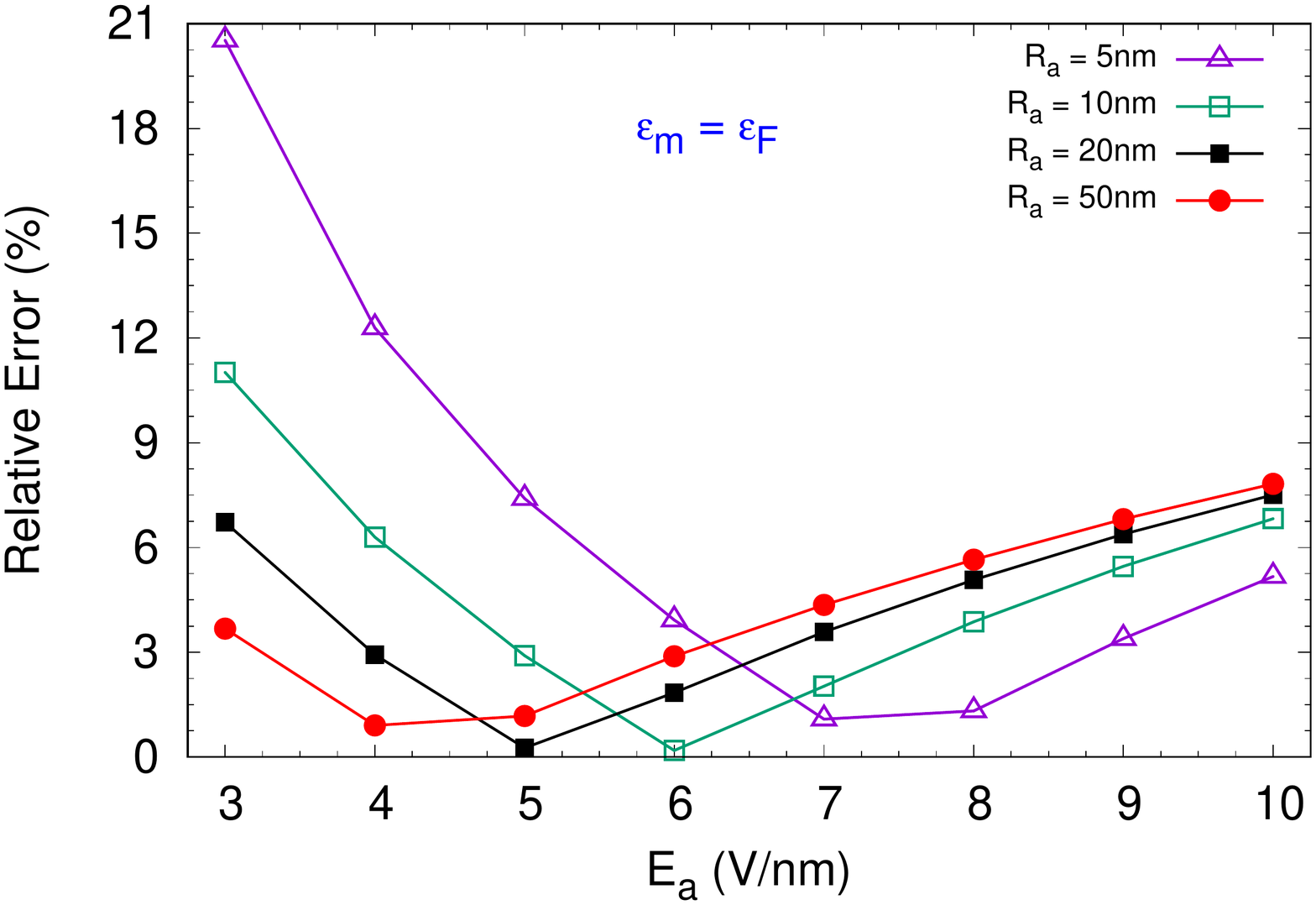}
\vskip -0.75 cm
\caption{The magnitude of the relative error in the analytical expression for
  the curvature-corrected current (Eq.~\ref{eq:anI}) compared to the exact WKB result for $\ce_m = \ce_F$.}
\label{fig:D0D1WKBEf}
\end{figure}

In Fig.~\ref{fig:D0D1WKB}, we compare the magnitude of the relative error in the net
current as given by Eq.~(\ref{eq:anI}) and (\ref{eq:gfact}) with respect to the exact WKB result
which has been used as the benchmark throughout this study with $\ce_m$ as the exact mean normal energy.
Clearly, the analytical expression
is adequate for a wide range of fields and apex radius of curvature. The increase in error at higher
fields is due to the linearization of $\cb_\tcc$ and $t_\tcm^{-2}$ in the variable $u$ which is a
measure of the distance from the apex. This is however a small price to pay for a compact
analytical expression for the net emission current.

For the sake of comparison, we also show the relative errors in the net current obtained using
the analytical expressions in 
Eq.~(\ref{eq:anI}) and (\ref{eq:gfact}) with $\ce_m = \ce_F$. The trends are similar to those
shown in section \ref{sec:num} where the linearized current density is integrated numerically
over the emitter end-cap. The errors are more pronounced at smaller apex radius of curvature
and apex field strengths. Clearly, linearization at the mean normal energy ensures
smaller errors over a wide range of fields and radius of curvature.

\begin{figure}[thb]
\vskip -0.5 cm
%\centering
\hspace*{-1.05cm}\includegraphics[width=0.62\textwidth]{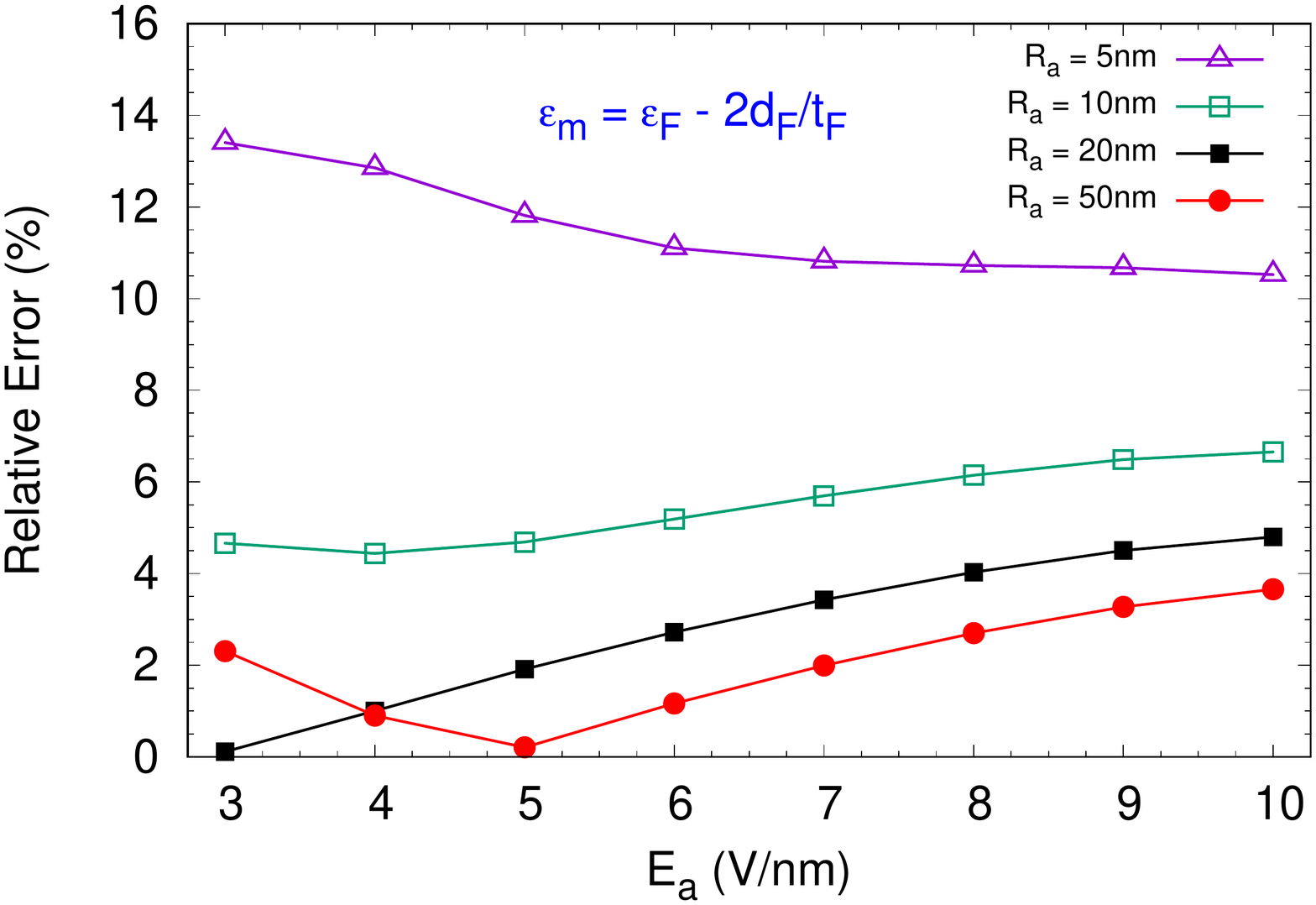}
\vskip -0.75 cm
\caption{The magnitude of the relative error in the analytical expression for
  the curvature-corrected current (Eq.~\ref{eq:anI}) compared to  the exact WKB result.
  Here, $\ce_m = \ce_F - 2d_F/t_F$.}
\label{fig:D0D1WKB_approx}
\end{figure}

While the errors are reasonably small when the exact mean normal energy is used, it
require the computation of integrals that marginally offsets the use of an analytical expression
for the net current.
In Fig.~\ref{fig:D0D1WKB_approx}, we provide a comparison
of the magnitude of relative errors with respect to the exact WKB result, using the approximate
$\ce_m = \ce_F - 2d_F/t_F$. While the errors for $R_a = 5$nm are somewhat large, the
approximate value of the mean normal energy may be used profitably for $R_a \geq 10$nm.

\begin{table}[htb]
 \begin{center}
   \caption{Comparison of time required for $10000$ evaluations of the net emission current for $E_a \in [3,10]$V/nm and $R_a \in [5,50]$nm. The `Scale Factor' is the ratio of the time taken by  `WKB Exact' and the time taken by a given method. It gives a rough indication of the speed-up achieved. Here $\ce_m$ is the
   mean normal energy. Also shown is the average relative error with respect to `WKB exact'.}
   \vskip 0.25 in
    \label{tab:time}
    \begin{tabular}{|l|c|c|c|}
      \hline
      Method & Time (s) & Scale Factor & Average Error \\  \hline \hline    
      WKB Exact & 315.1 s & 1 & --- \\ \hline
      WKB Fit & 23.8 s & 13.24 & 1.67\% \\ \hline
      Eq.~(\ref{eq:anI}) with  & 56.7 s & 5.55 & 2.06\% \\ 
      $\ce_m$ exact & & & \\ \hline
      Eq.~(\ref{eq:anI}) with  & 0.0003 s & $10^6$ & 3.14\% \\
      $\ce_m$ approximate & & & \\ 
      \hline
    \end{tabular}
\end{center} 
\end{table}

Finally, Table \ref{tab:time} provides a  comparison of the CPU time (in seconds) required on a standard desktop to serially compute the net emission current for $10^4$ combinations of $E_a$ and $R_a$ in the 
range of apex fields and radius of curvature considered in this paper. Thus, there are 100 values of $R_a$ spaced
uniformly in the range [5,50]nm and 100 values of $E_a$ spaced uniformly in the [3,10]V/nm range.
In the table, `WKB Fit' refers to the use of Eq.~(\ref{eq:G3}) for the Gamow factor and numerical integration over energy
 while `WKB exact' refers to the `exact' numerical evaluation of the Gamow factor followed by numerical integration
 over energy. The last two rows refer to the analytical formula for the net current of Eq.~(\ref{eq:anI}) with
 `$\ce_m$ exact' evaluated as outlined in appendix \ref{sec:appenMEAN} and
 `$\ce_m$ approximate' as  $\ce_m \approx \ce_F - 2d_F/t_F$. Clearly, linearization at the approximate mean normal
 energy results in fast computation of the net emission current using Eq.~(\ref{eq:anI})
 by a factor $\approx 80000$ compared to `WKB fit' and about $10^6$
  compared to the `WKB exact' result. This is only marginally offset by a larger error
  for $R_a = 5$nm as seen in Fig.~\ref{fig:D0D1WKB_approx}. The average relative error in the
  $5-50$nm range is however small as shown in Table \ref{tab:time}.

\section{Conclusions}

We have presented an expression for the curvature-corrected current density obtained by
linearization at an energy $\ce_m \leq \ce_F$ and insertion of a correction term to account for
the Kemble transmission coefficient. Numerical results show that the mean normal energy is a suitable
candidate for the linearization energy $\ce_m$ and predicts the net emission current
to within 3\% accuracy compared to the exact WKB result for $R_a \geq 5$nm and over
a wide range of field.

We have also obtained an analytical expression for the net emission current using the
generalized cosine law of local field variation. It requires only the apex radius
of curvature $R_a$ and the apex electric field $E_a$ and is able to calculate the
net field-emission current to within 6\% accuracy compared to the current obtained
by explicitly integrating the exact WKB current density over the emitter tip
for $R_a \geq 5$nm and a wide range of apex fields.

Both of these results are expected to be useful in dealing with sharp emitters having 
tip radius $R_a \geq 5$nm. The expression for current density can be used in all situations
including those where the emitter does not have any special symmetry. 
On the other hand, the expression for the net emission current is extremely useful
for axially symmetric emitters with smooth locally parabolic tips mounted in a
parallel plate configuration, considering that
the speed-up achieved in current computation is enormous.
The accuracies obtained in all cases are good, given that even minor experimental uncertainties can
lead to far larger changes in the net emission current. 

Finally, the analytical expression for emission current is especially useful
for a fast determination of net emission current from a large area field emitter have thousands of
axially symmetric emitters \cite{db_rudra_2018,db_rudra_2020,db_hybrid}.

%\section{Acknowledgements} 

\section{Author Declarations}

\subsection{Conflict of interest} There is no conflict of interest to disclose.
\subsection{Data Availability} The data that supports the findings of this study are available within the article.

\subsection{Author Contributions} {\bf Debabrata Biswas} Conceptualization (lead), data curation (equal), formal analysis (equal), methodology (lead), software (equal), validation (supporting), visualization (equal), original draft (lead), review and editing (supporting). \\

{\bf Rajasree Ramachandran} Conceptualization (supporting), data curation (equal), formal analysis (equal), methodology (supporting), software (equal), validation (lead), visualization (equal), original draft (supporting), review and editing (lead).

%\vskip -0.25 in

\appendix
\section{The tunneling potential}
\label{appendixA}

The electric field, $E_l$, close to the emitter surface is assumed to be a constant so that
the corresponding potential can be expressed as $V_{ext}(s) = -E_l s$ where $s$ is the normal
distance from the surface of the emitter and $E_l$ is the magnitude of the local electric field.
The assumption holds good when the radius of
curvature at the emission site is large (typically $R > 100$nm).

As $R$ decreases, corrections become important and these can be expressed as powers of
$s/R$. Thus,

\be
V_{ext}(s) = -E_l s \left[ 1 + c_1 \frac{s}{R} + \ldots + c_n \left(\frac{s}{R}\right)^n
  + \ldots \right].
\ee

\noi
The effective spherical approximation
used in Ref.~[\onlinecite{ludwick2021}] leads to $c_n = (-1)^n$ so that
$V_{ext}(s) = -E_l R \left[1 - 1/(1 + s/R)\right]$ with $R^{-1} = 2/(R_1^{-1} + R_2^{-1})$.
In Ref.~[\onlinecite{db_rr_gs_2018}], following the analysis of the hemiellipsoid, the hyperboloid
and the hemisphere, it was concluded that $c_1 = -1$, $c_2 = 4/3$ and $R \approx R_2$
where $R_2$ is the second (smaller) principle radius of curvature. With these identifications, the
external potential was found to approximate the numerically determined external potentials
for other emitter shapes as well\cite{db_rr_gs_2018}. Ref.~[\onlinecite{rr_db_2021}] uses the
nonlinear line charge model\cite{db_gs_rk_2016} for axially symmetric emitters to arrive at an approximate 
form close to the apex. In the following, we shall show that the results of both Ref.~[\onlinecite{db_rr_gs_2018,rr_db_2021}]
can be recast in the form where \{$c_1 = -1$, $R = R_m$\} exactly while \{$c_2 = 4/3$, $R^2 = R_m^2$\}
is approximate but fairly accurate close to the apex.

In addition to the approximate results in section II\cite{db_rr_gs_2018},
Ref.~[\onlinecite{db_rr_gs_2018}] also provides in the appendix,
a systematic expansion of the external potential in powers of $s$ for the hemi-ellipsoid.
In terms of the prolate spheroidal co-ordinates ($\eta,\xi,\varphi)$

\bea
x & = & L \sqrt{(\eta^2 - 1)(1 - \xi^2)} \cos\varphi \\
y & = & L \sqrt{(\eta^2 - 1)(1 - \xi^2)} \sin\varphi \\
z & = & L \eta \xi
\eea

\noi
it was found that

\be
V_{ext}(s) = V(s) =  \left[ d_1 s + d_2 s^2 + d_3 s^3 \right]
\ee

\noi
where

\bea
d_1 & = & V_{\eta} a_1 \\
d_2 & = & V_{\eta} a_2 + \frac{1}{2} V_{\eta \eta} a_1^2 \\
d_3 & = & V_{\eta} a_3 + V_{\eta \eta} a_1 a_2 + V_{\xi \eta} a_1 b_2 + \frac{1}{6} V_{\eta \eta \eta} a_1^3.
\eea

\noi
The derivatives of the potential at a point ($\eta_0,\xi_0$) on the surface of the hemiellipsoid
are

\bea
V_{\eta} &  =  & E_l h_\eta \\
V_{\eta \eta} & = & -E_l h\eta \frac{2\eta_0}{\eta_0^2 -1} \\
V_{\eta \xi} & = & E_l h_\eta \frac{1}{\xi_0} \\
V_{\eta \eta \eta} & = & E_l h_\eta \frac{8\eta_0^2}{(\eta_0^2 - 1)^2}
\eea

\noi
where

\bea
h_\eta &  = & L \sqrt{(\eta_0^2 - \xi_0^2)/(\eta_0^2 - 1)} \\
h_\xi  & = &  L \sqrt{(\eta_0^2 - \xi_0^2)/(1 - \xi_0^2)}.
\eea

\noi
The coefficients

\bea
a_1 & = & \frac{1}{h_\eta} \\
a_2 & = & \frac{1}{2h_{\xi}^2} \frac{\eta_0}{\eta_0^2 - \xi_0^2} \\
a_3 & = & - \frac{1}{2 h_\eta h_{\xi}^2} \frac{\eta_0^2 + \xi_0^2}{(\eta_0^2 - \xi_0^2)^2} \\
b_2 & = & - \frac{1}{2h_{\xi}^2} \frac{\xi_0}{\eta_0^2 - \xi_0^2}
\eea

\noi
while the principle radii of curvature are

\bea
R_1  & = & R_a \frac{(\eta_0^2 - \xi_0^2)^{3/2}}{(\eta_0^2  - 1)^{3/2}} \\
R_2  & = & R_a \frac{(\eta_0^2 - \xi_0^2)^{1/2}}{(\eta_0^2  - 1)^{1/2}}
\eea

\noi
where $R_a$ is the apex radius of curvature. On putting together these results, the values of $d_1, d_2$ and $d_3$ are

\bea
d_1 & = & - E_l \\
d_2 & = &  \frac{ E_l}{2R_a} \frac{(\eta_0^2 - 1)^{1/2}}{(\eta_0^2 - \xi_0^2)^{3/2}} \left[ 2\eta_0^2 - 1 - \xi_0^2 \right] \nonumber\\
& = &  \frac{E_l}{R_m} \\
d_3 & = & -\frac{4}{3} \frac{E_l}{R_m^2}  \frac{\left[3 + 4\eta_0^4 + \xi_0^4 - 2\eta_0^2(3 + \xi_0^2)\right]}{(1 + \xi_0^2 - 2\eta_0^2)^2} \\
& = & -\frac{4}{3} \frac{E_l}{R_m^2} \left[ 1 - 2\frac{(\eta_0^2 - 1)(1 - \xi_0^2)}{(1 + \xi_0^2 - 2\eta_0^2)^2} \right] \\
& = & -\frac{4}{3} \frac{E_l}{R_m^2} \left[ 1 - \cal{C} \right]
\eea

\noi
where

\bea
R_m & = & \frac{2}{(1/R_1 + 1/R_2)} \nonumber \\
& = & 2R_a \frac{\eta_0^2 - \xi_0^2)^{3/2}}{(\eta_0^2 - 1)^{1/2}(2\eta_0^2 - 1 - \xi_0^2)}.
\eea

\noi
The external potential thus takes the form

\be
V_{ext}(s) = -E_l s \left[ 1 - \left(\frac{s}{R_m}\right) + \frac{4}{3} \left(\frac{s}{R_m}\right)^2 \left( 1 - {\cal{C}} \right) \right]
\ee

\noi
for the hemiellipsoid emitter. In terms of $\rho_0^2 = x_0^2 + y_0^2$ where $x_0,y_0$ are on the surface of the hemiellipsoid,
the correction term ${\cal{C}} = \rho_0^2/(2R_a^2)$. Thus,

\be
V_{ext}(s) = -E_l s \left[ 1 - \frac{s}{R_m} + \frac{4}{3} \frac{s^2}{R_m^2} \left( 1 - \frac{1}{2} \frac{\rho_0^2}{R_a^2} \right) \right]. \label{eq:Vext_he}
\ee

\noi
Note that close to the apex, $\rho/R_a << 1$ while $R_m \approx R_2$.

A more general result, valid for all axially symmetric emitters in a parallel plate geometry, was arrived at using the
nonlinear line charge model\cite{rr_db_2021}. In such cases, the external potential can be expressed as (see Eq. (35) of Ref [\onlinecite{rr_db_2021}]), 

\be
V_{ext}(s) \approx -E_l s \left[1 - \frac{s}{R_a}(1 - \frac{\rho_0^2}{R_a^2}) + \frac{4}{3}\frac{s^2}{R_a^2}(1 - \frac{5}{2}\frac{\rho_0^2}{R_a^2})  \right]. \label{eq:35}
\ee

\noi
Close to the apex $1/R_m \approx (1/R_a)(1 - \rho_0^2/R_a^2)$ while $1/R_m^2 \approx (1/R_a^2)(1 - 2\rho_0^2/R_a^2)$. Thus, Eq.~(\ref{eq:35}) can be expressed as

\be
V_{ext}(s) \approx -E_l s \left[1 - \frac{s}{R_m} + \frac{4}{3}\frac{s^2}{R_m^2}(1 - \frac{1}{2}\frac{\rho_0^2}{R_a^2})  \right].
\ee

\noi
This is identical to the result obtained for the hemiellipsoid (see Eq.~\ref{eq:Vext_he}) but applicable
generally for all axially symmetric emitters. Approximating $(1 - \rho_0^2/(2R_a^2)) \approx 1$
leads us to an approximate universal form for the external potential (see Eq.~(\ref{eq:ext_pot})) close to the emitter surface.

Note that Eq.~(\ref{eq:35}) can also be expressed as

\be
V_{ext}(s) \approx -E_ls \left[ 1 - \frac{s}{R_m} + \frac{4}{3} \frac{s^2}{R_1 R_m} \right]
\ee

\noi
The correction terms, $\frac{s}{R_m}$ and $\frac{4}{3} \frac{s^2}{R_1 R_m}$ are exact for any point 
on the hemiellipsoid surface. For other emitter shapes\cite{rr_db_2021},
the two correction terms may have extra factors
that can be ascribed to the non-linear line charge distribution. Since these have been ignored
as an approximation, we choose to adopt the form in Eq.~(\ref{eq:ext_pot}) with $R = R_m$ as an approximate but
accurate representation of the external potential in the tunneling region.

Finally, while the change from $R_2$ to $R_m$ reduces the need for approximations, its impact on the
net field emission current is small
compared to a neglect of the second correction term $4s^2/(3R_m^2)$, especially at smaller values of $R_a$ and $E_a$.
For instance, at $R_a = 5$nm and workfunction $\phi = 4.5$eV, the error in net emission
current on using $R_2$ in Eq.~(\ref{eq:ext_pot})
is about 13\% at $E_a = 5$V/nm, while it is around 62\% on ignoring $4s^2/(3R_m^2)$ altogether.
At $E_a = 4$V/nm, the error in net emission current on using $R_2$ in Eq.~(\ref{eq:ext_pot}) remains
roughly the same while the error grows to around 92\% on ignoring $4s^2/(3R_m^2)$.
In each of these cases, the exact WKB method is used and the benchmark
current is obtained using $R_m$ in Eq.~(\ref{eq:ext_pot}).

%\appendix*
\section{The coefficients $\bf{D_0,D_1,F_0}$ and $\bf{F_1}$}
\label{sec:appen}

We shall briefly outline the derivation of the coefficients $D_0, D_1, F_0$ and $F_1$
and state the results. The dependence on $u$ in $\cb_\tcc$ and $t_\tcm$ arise from the
variation in $E_l$ and $x_m$ over the surface of the emitter. Thus,  

\be
\cb_\tcc = \cb_\tcc(E_l,x_m) = \cb_\tcc(\frac{E_a}{1 + u},\frac{x_\tma}{1 + u}) \approx D_0 + D_1 u
\ee

\noi
so that $D_0 = \cb_\tcc(E_a,x_\tma)$. In the above, $x_m = \varphi_m/(qE_l R_m) \approx x_\tma/(1 + u)$ where
$x_\tma = \varphi_m/(q E_a R_a)$. The approximation holds for tall emitters where $E_l = E_a (z/h)/\sqrt{(z/h)^2 + (\rho/R_a)^2} \approx E_a/\sqrt{1 + \rho^2/R_a^2}$. The coefficient $D_1$ can be written as

\bea
D_1 & = & \left( \frac{d \cb_\tcc}{du} \right)_{{u=0}} = \left(\frac{dE_l}{du} \right)_{{u=0}} \times \left( \frac{d \cb_\tcc}{dE_l}\right)_{E_l = E_a}  \nonumber \\
&  + &  \left(\frac{dx_m}{du} \right)_{{u=0}} \times \left( \frac{d \cb_\tcc}{dx_m}\right)_{x_m = x_\tma}.
\eea

\noi
Since $E_l = E_a/(1 + u)$, $(dE_l/du)_{u=0} = -E_a$. Similarly, since $x_m = x_\tma/(1 + u)$,
$(dx_m/du)_{u=0} = -x_\tma$. Thus,

\be
D_1 =  -E_a (d\cb_\tcc/dE_l)_{E_l = E_a} - x_\tma (d\cb_\tcc/dx_m)_{x_m = x_\tma}
\ee

\noi
This can be further expressed as

\be
\begin{split}
  D_1 & =  \cb_\tcc(E_a) - \frac{\Bfn \varphi_m^{3/2}}{E_a}\left[E_a \frac{d\nu_\tcm}{dE_l}\Big|_{E_a} + x_\tma \frac{d\nu_\tcm}{dx_m}\Big|_{x_\tma} \right] \\
  & + \frac{g\varphi_m^{1/2} (\ce_F - \ce_m)}{E_a} \left[E_a\frac{dt_\tcm}{dE_l}\Big|_{E_a} + x_\tma \frac{dt_\tcm}{dx_m}\Big|_{x_\tma} \right].
\end{split}
\ee

\noi
It is simpler to express $d\nu_\tcm/dE_l$ and $dt_\tcm/dE_l$ as

\bea
\frac{d\nu_\tcm}{dE_l}\Big|_{E_a} & = & \frac{d\nu_\tcm}{d(y^2)}\Big|_{y_\tma}  \frac{d(y^2)}{dE_l}\Big|_{E_a} \\
\frac{dt_\tcm}{dE_l}\Big|_{E_a} & = & \frac{d t_\tcm}{d(y^2)}\Big|_{y_\tma}  \frac{d(y^2)}{dE_l}\Big|_{E_a}
\eea

\noi
and use the fact that $d(y^2)/dE_l = 4qB/\varphi_m^2$. The quantities $\frac{d\nu_\tcm}{dx_m}\Big|_{x_\tma}$
and $\frac{dt_\tcm}{dE_l}\Big|_{x_\tma}$ can be obtained directly and expressed as

\bea
\frac{d\nu_\tcm}{dx_m}\Big|_{x_\tma} & = & w_1(y_\tma) + 2 x_\tma w_2(y_\tma) + 3 x_\tma^2 w_3(y_\tma) \nonumber \\
\frac{dt_\tcm}{dx_m}\Big|_{x_\tma} & = & t_1(y_\tma) + 2 x_\tma t_2(y_\tma) + 3 x_\tma^2 t_3(y_\tma) \nonumber .
\eea

\noi
These results can be combined to obtain

\be
\begin{split}
  & D_1  =  \cb_\tcc(E_a) \\
  &  - \Bfn \frac{4qB}{\varphi_m^{1/2}} \frac{d\nu_\tcm}{d(y^2)}\Big|_{y_\tma^2} +
   \frac{4Bqg (\ce_F - \ce_m)}{\varphi_m^{3/2}}\frac{dt_\tcm}{d(y^2)}\Big|_{y_\tma^2}\\
& -\frac{\Bfn \varphi_m^{3/2}}{E_a} x_\tma \frac{d\nu_\tcm}{dx_m}\Big|_{x_\tma} + \frac{(\ce_F - \ce_m)}{d_\tma} x_\tma \frac{dt_\tcm}{dx_m}\Big|_{x_\tma} \label{eq:d1}
\end{split}
\ee

\noi
where $y_{\tma}^2 = 4qBE_a/\varphi_m^2$. Finally,

\be
\begin{split}
 ~~~ &  \left(\frac{d\nu_\tcm}{d(y^2)}\right)_{y^2 = y_\tma^2}  =  \\
%  & - \frac{1}{E_a}\left[ x_\tma w_1(y_\tma) + x_\tma^2 w_2(y_\tma) + x_\tma^3 w_3(y_\tma)\right] + \\
&  u_0(y_\tma) + x_\tma u_1(y_\tma) + x_\tma^2 u_2(y_\tma) + x_\tma^3 u_3(y_\tma)
\end{split}
\ee

\noi
with

\bea
u_0(y_\tma) & = & -1 + \frac{1}{6}(1 + \ln y_\tma^2) + \frac{1}{6}  \\
u_1(y_\tma) & = & -\frac{2}{11} + \frac{2}{80}y_\tma^2 + \frac{1}{400}(1 + \ln y_\tma^2) \\
u_2(y_\tma) & = & \frac{2}{11} + \frac{2}{3}y_\tma^2 + \frac{1}{400}(1 + \ln y_\tma^2) \\
u_3(y_\tma) & = & \frac{39}{20} + \frac{2}{3}y_\tma^2 - \frac{1}{300}(1 + \ln y_\tma^2).
\eea

\noi
Similarly,

\be
\begin{split}
 ~~~ &  \left(\frac{d t_\tcm}{d(y^2)}\right)_{y^2 = y_\tma^2}  =  \\
%  & - \frac{1}{E_a}\left[ x_\tma t_1(y_\tma) + x_\tma^2 t_2(y_\tma) + x_\tma^3 t_3(y_\tma)\right] + \\
&  p_0(y_\tma) + x_\tma p_1(y_\tma) + x_\tma^2 p_2(y_\tma) + x_\tma^3 p_3(y_\tma)
\end{split}
\ee

\noi
with

\bea
p_0(y) & = & \frac{1}{9} - \frac{1}{18}(1 + \ln y^2) + \frac{1}{6}  \\
p_1(y) & = & -\frac{237}{1100} - \frac{1}{240}y^2 - \frac{7}{2400}(1 + \ln y^2) \\
p_2(y) & = & \frac{589}{3300} + \frac{1}{9}y^2 + \frac{1}{400}(1 + \ln y^2) \\
p_3(y) & = & \frac{2929}{900} + \frac{2}{9}y^2 - \frac{1}{45}(1 + \ln y^2).
\eea

\noi
This completes the evaluation of $D_1$ in Eq.~(\ref{eq:d1}).

The coefficients $F_0$ and $F_1$ are defined as

\be
\frac{1}{t_\tcm^2(E_l,x_m)} = \frac{1}{t_\tcm^2(\frac{E_a}{1+u},\frac{x_\tma}{1+u})} \approx F_0 + F_1 u
\ee

\noi
so that $F_0 = 1/t_\tcm^2(E_a,x_\tma)$. The coefficient $F_1$ is

\be
F_1 = \frac{d t_\tcm^{-2}}{du} \Big|_{{u=0}} =  \frac{2}{t_\tcm^3} \left[E_a \frac{d t_\tcm}{dE_l}\Big|_{E_a}  +
  x_\tma \frac{d t_\tcm}{dx_m}\Big|_{x_\tma} \right]
\ee

\noi
which can be finally  expressed as

\be
%\begin{split}
F_1 = \frac{2}{t_\tcm^3} \left[ \frac{4qBE_a}{\varphi_m^2} \frac{dt_\tcm}{d(y^2)}\Big|_{y_\tma^2} + x_\tma \frac{d t_\tcm}{dx_m}\Big|_{x_\tma} \right].
\ee

\section{The `exact' mean normal energy}
\label{sec:appenMEAN}

The exact mean normal energy can be determined starting with the joint distribution $f(\ce_N,\tth)$ or equivalently
$f(\ce_N,\rho)$ \cite{db_dist}. In terms of $\rho$, it can be expressed as $\langle \ce_N \rangle = S_1/S_2$
where

\bea
S_1 & = & \int \int d\rho d\ce_N~ \rho \sqrt{1 + \rho^2/R_a^2} (\ce_F - \ce_N) \ce_N T(\ce_N,\rho) \nonumber \\
S_2 & = & \int \int d\rho d\ce_N~ \rho \sqrt{1 + \rho^2/R_a^2} (\ce_F - \ce_N) T(\ce_N,\rho). \nonumber 
\eea

\noi
In the above $T(\ce_N,\rho) \approx 1/(1 + e^{G(\ce_N,\rho)})$ is the transmission coefficient for an electron having normal energy $\ce_N$
at a point $\rho$ on the emitter-tip $z \approx h - \rho^2/(2R_a)$, having a local field $E_l = E_a (z/h)/\sqrt{z^2/h^2 + \rho^2/R_a^2}$.
It can be determined using Eqns.~(\ref{eq:G3}) - (\ref{eq:ws}) for the Gamow factor $G$.


\begin{thebibliography}{99}
\bibitem{assumed_constant} The emission area was assumed to be independent of the local field.
\bibitem{see1} For a recent analysis on a single emitter not requiring curvature correction, see [\onlinecite{popov23}].
\bibitem{high_field} At high fields, contributions beyond $\rho \approx R_a$ cannot be
  altogether neglected. While the validity of the parabolic approximation and the cosine law (except in
  hemi-ellipsoids) start breaking down for $\rho > R_a$, the curvature corrected
  current density of Eq.~(\ref{eq:MGshift}) continues to hold and can be used
  to determine the net emission current.
\end{thebibliography}
\end{document}